\documentclass[prd,aps,nofootinbib,superscriptaddress,floatfix,preprintnumbers,twocolumn]{revtex4}%showpacs

\usepackage[colorlinks, citecolor=blue,anchorcolor=blue,menucolor=blue, linkcolor=blue,filecolor=blue,runcolor=blue,urlcolor=blue,frenchlinks=true]{hyperref}
\usepackage{amsmath,amsfonts,amssymb,mathtools,cases,slashed,bm}
\usepackage{epsfig, graphicx}
\usepackage{color}
\usepackage[dvipsnames]{xcolor}
\usepackage{tabularx}
\usepackage{physics}
\usepackage{hhline}
\usepackage{mathrsfs}
\usepackage[normalem]{ulem}
\usepackage{multirow}
\usepackage{float}
\usepackage{appendix}

\newcommand{\UPDATE}[1]{#1}%\textcolor{red}{#1}}

\begin{document}

\title{First Nucleon Gluon PDF from Large Momentum Effective Theory}

\author{William Good}
\email{goodwil9@msu.edu}
\affiliation{Department of Physics and Astronomy, Michigan State University, East Lansing, Michigan 48824, USA}
\affiliation{Department of Computational Mathematics,
  Science and Engineering, Michigan State University, East Lansing, Michigan 48824, USA}
\author{Fei Yao}
\email{fyao@bnl.gov}
\affiliation{Physics Department, Brookhaven National Laboratory, Upton, New York 11973, USA}

\author{Huey-Wen Lin}
\affiliation{Department of Physics and Astronomy, Michigan State University, East Lansing, Michigan 48824, USA}
%https://orcid.org/my-orcid?orcid=0000-0001-6281-944X

%%%%%%%%%%%%%%%%%%%%%%%%%%%%%%%%%%%%%%%%%%%%%%%%%%%%%%%%%%%%%%%%%%%%%%%%%%%
\begin{abstract}
We report the first nucleon gluon parton distribution function (PDF) using Large-Momentum Effective Theory (LaMET).
We focus on the gluon operator which was demonstrated to have the best signal-to-noise in the previous attempt~\cite{Good:2024iur} in computing gluon PDFs using LaMET.
We compute the corresponding Wilson coefficients needed for the hybrid-renormalized matrix elements and the matching kernel to convert the quasi-PDF to the lightcone one at the one-loop level.
We demonstrate that with the proper Wilson coefficients in place, the counterterms for the renormalization are independent of the hadron and mass within statistical error.
Using the resulting renormalization, we then compute the nucleon PDF using a HISQ ensemble generated by the MILC collaboration with $N_f=2+1+1$, $a \approx 0.12$~fm, with valence pion masses of 310 and 690 MeV and two gauge link smearing techniques.
Despite the physics effects of the heavier than physical pion masses and gauge link smearing, this calculation provides excellent proof of principle and compares reasonably with selected global fit results. 
\end{abstract}

%%%%%%%%%%%%%%%%%%%%%%%%%%%%%%%%%%%%%%%%%%%%%%%%%%%%%%%%%%%%%%%%%%%%%%%%%%%
\preprint{MSUHEP-25-024} %,BNL?}
\date{\today}

\maketitle

%%%%%%%%%%%%%%%%%%%%%%%%%%%%%%%%%%%%%%%%%%%%%%%%%%%%%%%%%%%%%%%%%%%%%%%%%%%

\section{Introduction}
\label{sec:intro}

%\begin{itemize}
%    \item Why study gluon PDFs
%    \item into to x-dependent PDFs
%    \item gluon PDFs studies to date
%    \item Our previous attempt in Ref.~\cite{Good:2024iur} https://arxiv.org/abs/2409.02750
%    \item this work summary
%\end{itemize}

Parton distribution functions (PDFs) play a fundamental role in high-energy particle physics, encoding the momentum distributions of quarks and gluons within protons and neutrons.
These functions are essential inputs for making precise predictions at hadron colliders such as the Large Hadron Collider (LHC), directly impacting the interpretation of a wide range of Standard Model (SM) and beyond-the-SM (BSM) processes.
Among the PDFs, the gluon distribution is particularly crucial due to the dominant role gluons play in the proton's momentum at high energies and in processes like Higgs-boson production and jet formation.
While quark PDFs are relatively well constrained by deep inelastic scattering and Drell-Yan data, the gluon PDF remains less precisely determined, especially in the intermediate and large momentum fraction ($x$) regions.
This uncertainty limits the precision of theoretical predictions for key observables at the LHC.
Improving the accuracy of the gluon PDF, therefore, has the potential to significantly enhance the sensitivity of searches for BSM physics and refine our understanding of the dynamics of quantum-chromodynamics (QCD) in hadronic collisions~\cite{Amoroso:2022eow}.

Lattice QCD (LQCD) provides a first-principles approach to studying the nonperturbative regime of the strong interaction by discretizing spacetime on a finite lattice.
Although LQCD was traditionally limited to the calculation of a few hadron structure observables such as moments of PDFs, recent breakthroughs have enabled LQCD calculations to access the Bjorken-$x$--dependence of PDFs.
Methods such as Large-Momentum Effective Theory (LaMET)~\cite{Ji:2013dva} and pseudo-PDFs~\cite{Radyushkin:2017cyf} are the most popular approaches to extracting lightcone correlations from lattice observables.
Over the past decade or so, substantial progress has been made not only in PDFs but also in other kinds of three-dimensional structures, such as generalized parton distributions (GPDs);
we refer interested readers to the reviews in Refs.~\cite{Ji:2020ect,Constantinou:2020hdm,Lin:2023kxn}.
However, the majority of lattice $x$-dependent results focus on the isovector quark distributions, and there are few works on the gluon PDFs.

The extraction of gluon PDFs in lattice QCD remains more challenging, due to their inherently noisier signals and the need to handle more complicated operator mixing.
Nevertheless, encouraging progress has been reported in recent years, mostly using the pseudo-PDF framework, obtaining the nucleon~\cite{Fan:2020cpa, Fan:2022kcb, HadStruc:2021wmh, Delmar:2023agv, HadStruc:2022yaw}, pion~\cite{Good:2023ecp}, and kaon~\cite{Salas-Chavira:2021wui} gluon PDFs.
However, the pseudo-PDF method often requires the use of phenomenology-inspired $x$-dependent PDF models to fit the ratios of lattice matrix elements; though, there are recent efforts to reduce model dependence in this methodology \cite{Karpie:2019eiq, Karpie:2021pap}.
LaMET, on the other hand, uses a Fourier transform of the lattice matrix elements to first obtain quasi-PDFs, which can be matched to the lightcone PDFs via a matching procedure accurate to powers of the inverse parton momentum.
Model dependence comes into play in LaMET in the extrapolation of large-distance matrix elements to obtain a reliable Fourier transform.
It would be of great community interest to obtain the gluon PDF through LaMET as well, so that we can achieve better control of the lattice systematics by comparing results from the two methodologies.

Recently, MSULat group aimed to advance the LQCD calculations of gluon PDFs using LaMET~\cite{Good:2024iur}.
We used high-statistics measurements at two pion masses (approximately 310 and 690~MeV) on a lattice ensemble with a spacing of about 0.12~fm.
The work explored the application of Coulomb gauge fixing to enhance signal quality at large Wilson-line separations, demonstrating its potential to improve the reliability of the calculations of gluon matrix elements.
By analyzing multiple gluon operators and comparing the resulting matrix elements for both nucleon and pion states, the study identified the most effective operator for extracting gluon PDFs.
%The results indicate that O(3) consistently produces better signal-to-noise renormalized matrix elements compared to O(1) and O(2).
%Despite these advancements, MSULat acknowledges several challenges that need to be addressed in future research.
%Achieving higher hadron boost momenta is necessary to implement hybrid-ratio renormalization reliably, which is crucial for accurate matching between quasi-PDFs and lightcone PDFs.
%Furthermore, the exploration of a more diverse set of gluon operators, along with their corresponding perturbative calculations, is essential to refine the extraction of gluon PDFs from lattice QCD.
However, without the perturbative QCD (pQCD) calculations needed for hybrid-ratio renormalization of this operator, we could only perform an analysis with an estimation of the renormalization.
%Overcoming these bottlenecks will enhance the precision of gluon PDF determinations, contributing to a deeper understanding of hadronic structure and informing experimental analyses in nuclear and high-energy physics.

%\FIXME{other paragraphs to fill in }

In this work, we took the lattice gluon matrix elements from our previous work~\cite{Good:2024iur} and calculated the hybrid-renormalization-scheme matching coefficients to obtain the first lightcone gluon PDF of the nucleon using the LaMET approach.
We report the next-to-leading-order (NLO) Wilson coefficients for the hybrid renormalization scheme for the gluon operator in Sec.~\ref{sec:analytic}.
We determine the counterterms in the hybrid renormalization and apply them to renormalize the LaMET matrix elements in Sec.~\ref{sec:numerical}.
We compare the renormalized matrix elements with those estimated from phenomenology, and present the first nucleon gluon PDF from LaMET, also compared with phenomenological fits.
We conclude and consider the future prospects for gluon PDFs in Sec.~\ref{sec:conclusion}.
%%%%%%%%%%%%%%%%%%%%%%%%%%%%%%%%%%%%%%%%%%%%%%%%%%%%%%%%%%%%%%%%%%%%%%%%%%%
\section{Formulation for LaMET and Hybrid Renormalization}
\label{sec:analytic}

The leading-twist gluon PDFs $g(x)$ of a hadron is defined as~\cite{Collins:1989gx},
\begin{align}\label{eq:LCPDF}
g(x,\mu) = \frac{1}{ 2 x P^+} \int_{-\infty}^\infty \frac{\mathrm{d} \xi^-}{2\pi} \,& e^{-i x P^+ \xi^-} \langle P | F_a^{+\mu}(\xi^-) \notag\\
&
 \times \mathcal{W}_{ab}(\xi^-, 0) \, F_{b,\mu}^+(0) | P \rangle,
\end{align}
where $|P\rangle$ denotes unpolarized hadron with momentum $P$ along the $z$-direction.  Here, $x$ is the longitudinal momentum fraction carried by the gluon, and $\mu$ denotes the renormalization scale in the $\overline{\rm MS}$ scheme. The light-front coordinates are given by $\xi^\pm=(\xi^t\pm \xi^z)/\sqrt{2}$ and gauge invariance is ensured by the Wilson line,
\begin{align}
\mathcal{W}_{ab}(\xi^-, 0)= \mathcal{P}\,{\rm exp}\big[ -i g \int_0^{\xi^-} \mathrm{d} \eta^- \, A_c^+(\eta^-) t^c \big]_{ab},
\end{align}
where the indices $\{a,b\}$ denote the color components in the adjoint representation of SU(3), and $t^c$ are the generators of the gauge group.	

According to LaMET, the gluon PDF can be extracted by calculating a spatially correlated matrix element:
\begin{align}
h^{\text{B}}(z,P_z)= \langle P | &F^{ti}(z)\mathcal{W}(z,0)F^t_{i}(0) \notag\\
&-F^{ij}(z)\mathcal{W}(z,0)F_{ij}(0)| P \rangle,
\end{align}
where $i$ and $j$ run over the transverse indices $\{x,y\}$, only.
Hereafter, we adopt the convention of omitting color indices for notational simplicity.

Although the local case does not strictly admit a multiplicative renormalization, we do not incorporate information from the local region, as the result is normalized and therefore insensitive to local contributions. The bare non-local quasi-light-front (LF) correlation can be multiplicatively renormalized as~\cite{Zhang:2018diq}
%contain z-dependent linear divergence brought by Wilson-line self-energy as well as z-independent logarithmic ultraviolet (UV) divergence
\begin{align}
h^{\text{B}}(z,P_z)=Z_L e^{-\bar{\delta} m |z|} \,h^{\text{R}}(z,P_z).
\end{align}
where $h^{\text{R}}(z,P_z)$ is the renormalized quasi-LF correlation that contains both $z$-independent logarithmic UV divergences, captured by $Z_L$, and $z$-dependent linear divergences from Wilson-line self-energy, denoted by $\bar{\delta} m$.
%To remove these divergences, several non-perturbative renormalization schemes have been proposed~\cite{Chen:2016fxx,Izubuchi:2018srq,Alexandrou:2017huk,Radyushkin:2018cvn,Braun:2018brg,Li:2018tpe}, but they often introduce unwanted non-perturbative contributions that distort the IR behavior.
%To avoid this, 
To remove these divergences, we adopt the hybrid renormalization scheme~\cite{Ji:2020brr}, treating short and long distances separately:
at short distances, we apply the ratio scheme~\cite{Radyushkin:2018cvn} using rest-frame matrix elements.
While at long distances, we use mass renormalization by fitting the zero-momentum matrix $h^{\text{B}}(z,0)$ to capture linear divergence effects.
The resulting renormalized quasi-LF correlation is expressed as:
\begin{equation}\label{eq:hybridscheme}
h^{\mathrm{R}}(z, P_z) =
\begin{cases}
\!\displaystyle\!
\frac{h^{\mathrm{B}}(0,0)}{h^{\mathrm{B}}(0,P_z)} \frac{h^{\mathrm{B}}(z,P_z)}{h^{\mathrm{B}}(z,0)} & z \leq z_s \\[12pt]
\!\displaystyle\!
\frac{h^{\mathrm{B}}(0,0)}{h^{\mathrm{B}}(0,P_z)}  \frac{h^{\mathrm{B}}(z,P_z)}{h^{\mathrm{B}}(z_s,0)} e^{(\delta m + m_0)(z - z_s)} & z > z_s
\end{cases}
\end{equation}
where $z_s$ separates short and long distances, the terms $\delta m$ and $m_0$ correspond to the mass renormalization and the renormalon ambiguity, respectively, and are introduced to cancel the linear divergence arising from the self-energy of the Wilson line. In the limit $z_s \to \infty$, the expression reduces to the standard ratio-scheme renormalization, which neglects the Wilson line self-energy contribution.
\UPDATE{Using the double ratio in the short distance is a good method for handling discretization errors associated with the non-commutativity of the $z \rightarrow 0$ and lattice spacing $a \rightarrow 0$ limits~\cite{Ji:2020brr}, though this scheme will cause the resulting gluon PDF to be normalized by its first moment $\langle x\rangle_g$.
We can see in a previous study of the continuum limit of the gluon PDF that discretization errors in the short distance using the ratio scheme are well under control within current statistical errors~\cite{Fan:2022kcb}.
It has been suggested that more control over discretization effects can be achieved using the self-renormalization method~\cite{LatticePartonLPC:2021gpi}, though this requires the use of data from multiple lattice spacings.}

\begin{figure*}[htbp]
\centering
        \includegraphics[width=.45\textwidth]{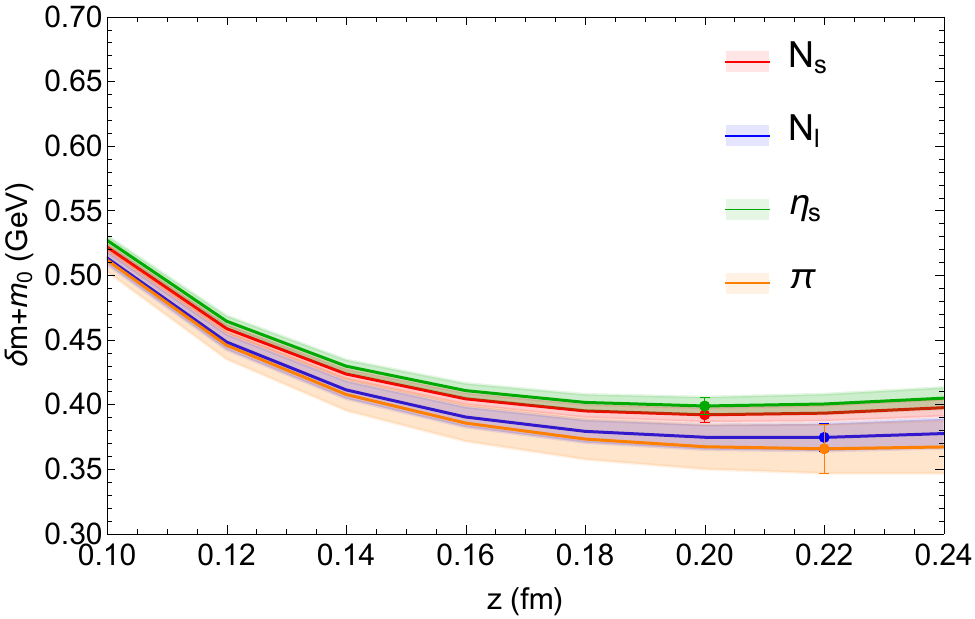}
\centering
        \includegraphics[width=.45\textwidth]{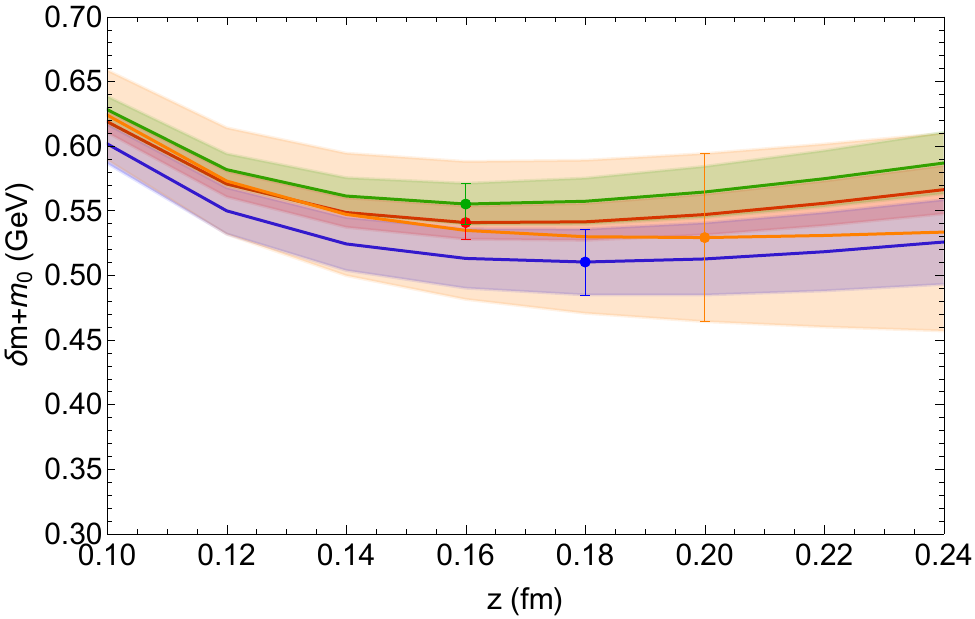}

    \caption{Fit results for the $\delta m + m_0$ at $\mu = 2 $ GeV for varying central point $z$ for each hadron for the two smearings, Wilson-3 (left) and HYP5 (right).
    The error bands give statistical fluctuations only.
    The points represent the choice of $\delta m + m_0$ that is used in the subsequent hybrid renormalization.}
    \label{fig:m0_vs_z}
\end{figure*}

%\FIXME{Wilson coefficients for  $\delta m + m_0$  } 
In particular, the availability of multiple lattice spacings allows for a separate determination of $\delta m$ and $m_0$ via fitting procedures~\cite{Ji:2020brr, LatticePartonLPC:2021gpi, Zhang:2023bxs}. 
However, in the case of a single lattice spacing, it is more practical to fit only their sum, $\delta m + m_0$, by matching the $P_z = 0$ bare matrix elements to the perturbatively calculated ``Wilson coefficients". The Wilson coefficient can be derived from Ref.~\cite{Balitsky:2021qsr}, gives
\begin{align}\label{eq:wil-coef}
\mathcal{H}(z,\mu) = 1 + \frac{\alpha_s C_A}{2\pi} \left( \frac{5}{6} \ln \left( \frac{z^2 \mu^2 e^{2\gamma_E}}{4} \right) + \frac{3}{2} \right).
\end{align}
Then we determine the combined parameter $\delta m + m_0$ in the short-distance region by fitting to the relation:
\begin{equation}\label{eq:dm_fit}
(\delta m + m_0) z  + I_0 \approx \ln\left[ \frac{\mathcal{H}(z,\mu)}{h^{\mathrm{B}}(z,0)} \right],
\end{equation}
where $I_0$ denotes a $z$-independent constant. 

%\FIXME{hybrid-renormalization matching kernel} 
The renormalized quasi-LF correlation function can be transformed into the so-called quasi-PDF via a Fourier transform. The gluon quasi-PDF is related to the light-front PDF through a factorization formula as following:
\begin{align}\label{eq:mtcheq}
 x\tilde{g}\left(x, P_z\right)= & \int_{-1}^1 \frac{d y}{|y|} C\left(\frac{x}{y}, \frac{\mu}{y P_z}\right)  y g(y, \mu)  +h.t. ,
\end{align}
where $g(y, \mu)$ is the light-front gluon PDF at the renormalization scale $\mu$, $C\left(x/y, \mu/(y P_z)\right)$ represents the perturbative matching coefficient, and $h.t.$ denotes higher-twist corrections suppressed by powers of $1/P_z$. 
\UPDATE{Notice that we neglect the contribution of mixing with quarks in this work, as it has been shown to have an effect smaller than $\sim10$\% in the pseudo-PDF studies~\cite{Fan:2022kcb}, which is well within statistical precision and smaller than the lattice systematic effects that we explore in this paper.} 
The perturbative matching coefficient in coordinate space up to the NLO in the ratio scheme can be found in Refs.~\cite{Balitsky:2019krf, Balitsky:2021qsr}. Then we obtain the corresponding momentum-space kernel directly via a Fourier transform, following the procedure outlined in Ref.~\cite{Yao:2022vtp}. The matching coefficients in the ratio and hybrid schemes are given as follows:

\begin{widetext}
\begin{equation}\label{eq:kernel}
C^{\text{ratio}}\left(\xi, \frac{\mu}{p_z}\right) 
= \delta(1-\xi) 
+ \frac{\alpha_s C_A}{2\pi} 
\left\{
\begin{array}{ll}
\left(
2 \frac{(1 - \xi + \xi^2)^2}{1 - \xi} \ln\left(\frac{\xi}{\xi-1}\right) 
+\frac{11 - 28\xi + 18\xi^2 - 12\xi^3}{6(1-\xi)}
\right)_{+(1)}, & \xi > 1 \\[1.5ex]
\left(
2 \frac{(1 - \xi + \xi^2)^2}{1 - \xi} \left[ -\ln(\frac{\mu^2}{4{p_z}^2}) + \ln\left[\xi(1-\xi)\right] \right]
- \frac{15 - 56\xi + 102\xi^2 - 96\xi^3 + 48\xi^4}{6(1-\xi)}
 \right)_{+(1)}, & 0 < \xi < 1 \\[1.5ex]
\left(
-2 \frac{(1 - \xi + \xi^2)^2}{1 - \xi} \ln\left(\frac{-\xi}{1-\xi}\right) 
- \frac{11 - 28\xi + 18\xi^2 - 12\xi^3}{6(1-\xi)} \right)_{+(1)}, & \xi < 0
\end{array}
\right.
\end{equation}

\begin{equation}
C^{\text{hyb.}}\left(\xi, \frac{\mu}{p_z}\right) =C^{\text{ratio}}\left(\xi, \frac{\mu}{p_z}\right) + \frac{\alpha_s C_A}{2\pi} \frac 5 6 \bigg[-\frac{1}{|1-\xi|} +\frac{2\text{Si}((1-\xi)z_s p_z)}{\pi(1-\xi)}\bigg]_+ . 
\end{equation}
\end{widetext}
where $\xi = \frac{x}{y}$ and $p_z$ denotes the momentum carried by the parton, which is given by $y P_z$.
%%%%%%%%%%%%%%%%%%%%%%%%%%%%%%%%%%%%%%%%%%%%%%%%%%%%%%%%%%%%%%%%%%%%%%%%%%%

\section{Numerical Results and Discussions}
\label{sec:numerical}

\begin{figure*}[htb]
    \centering
    \includegraphics[width=0.45\linewidth]{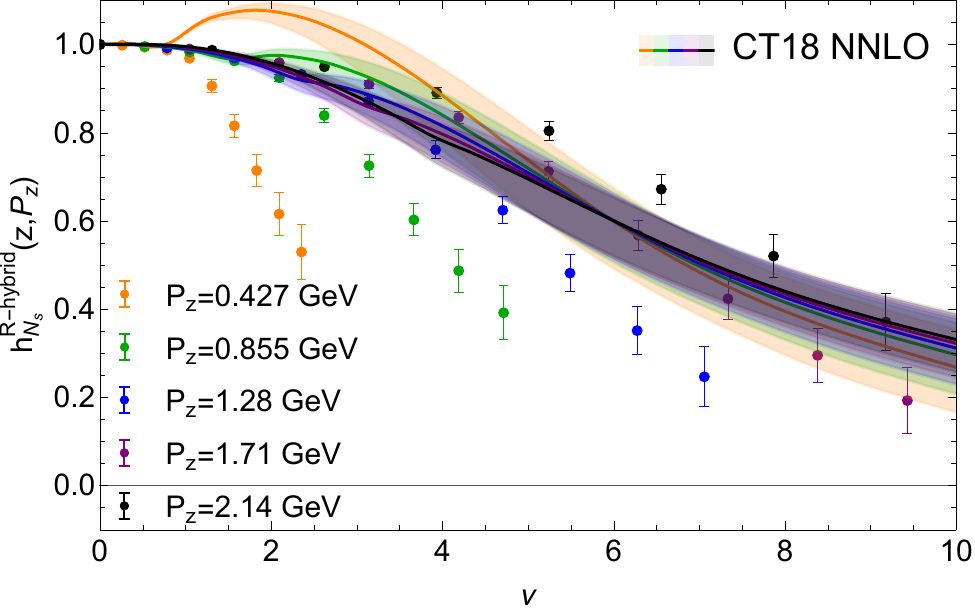}
    \includegraphics[width=0.45\linewidth]{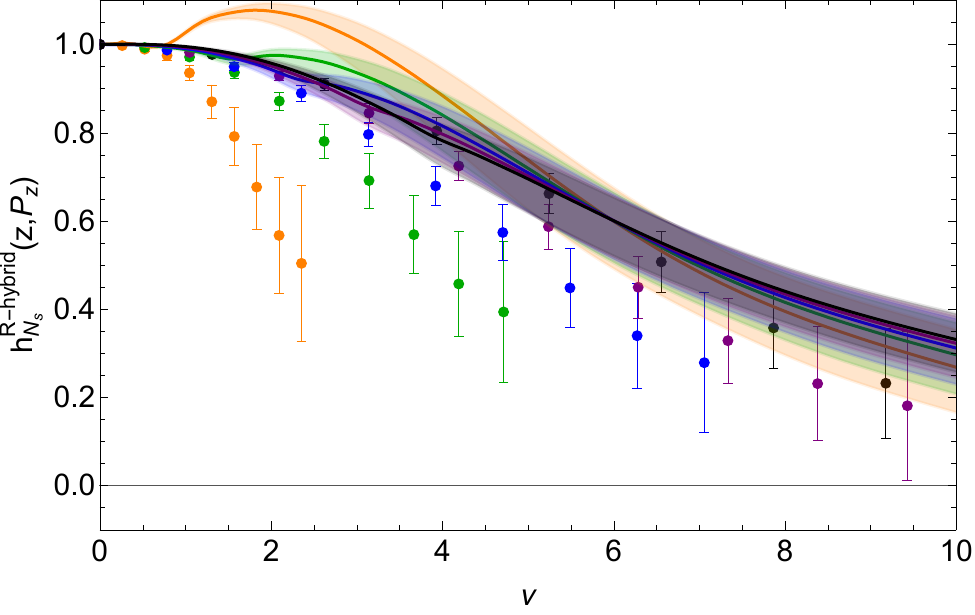}
    \includegraphics[width=0.45\linewidth]{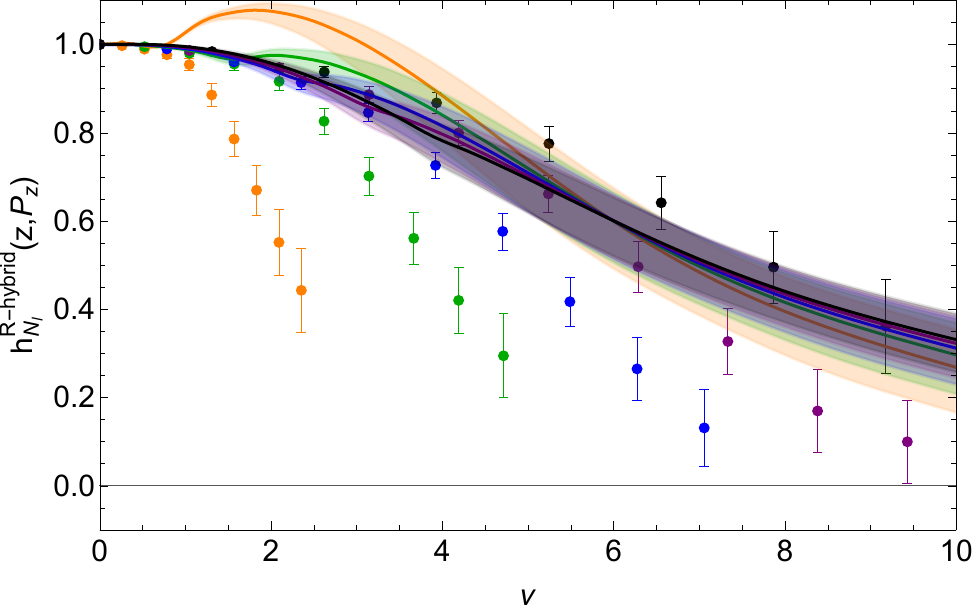}
    \includegraphics[width=0.45\linewidth]{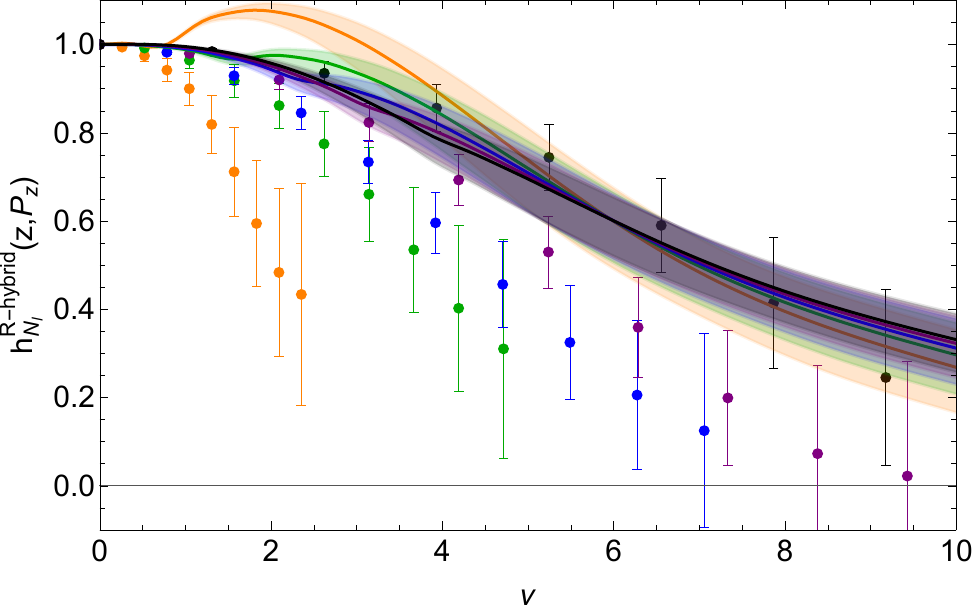}
    \caption{\UPDATE{Lattice (points) hybrid renormalized nucleon matrix elements with $z_s = 0.36$~fm for the Wilson-3 (left column) and HYP5 (right column) smearing for $M_\pi \approx 690$ (top row) and $310$ (bottom row) GeV compared to the matrix elements reconstructed from the CT18 gluon PDF~\cite{Hou:2019efy}, shown by the colored bands which have the same color scheme as the lattice data for differing momenta.}}
    \label{fig:hybrid-lat-pheno}
\end{figure*}

We use the high-statistics data from Ref.~\cite{Good:2024iur} on one ensemble with lattice spacing $a \approx 0.12$~fm at valence pion masses $M_{\pi} \approx 310$ and $690$~MeV generated by the MILC collaboration~\cite{MILC:2013znn} using 2+1+1 flavors of highly improved staggered quarks (HISQ)~\cite{Follana:2007rc} with the lattice volume of $24^3 \times 64$.
We use Wilson clover fermions in the valence sector and tune the valence quark mass to reproduce the light and strange mass in the HISQ sea.
We computed 1,296,640 two-point correlators across 1013 configurations for both the nucleon and pion at the light and heavy pion masses with fixed Gaussian momentum smearing~\cite{Bali:2016lva}.
The nucleons and pions are referred to as the light ($N_l$) and strange ($N_s$) nucleons, the pion ($\pi$), and $\eta$-strange ($\eta_s$) for the light and heavy pion masses, respectively.
We will focus mostly on the nucleons but we include the pion data to understand how the external states affect the calculation of $\delta m + m_0$.
We measured two sets of gluon loops with 5 steps of hypercubic \UPDATE{\cite{Hasenfratz:2001hp}} (HYP5) smearing and with Wilson flow\UPDATE{\cite{Luscher:2010iy}} with a flow time of $T=3 a^2$ (Wilson3) on the gauge links \UPDATE{in order to achieve reasonable signal.
A more detailed study of smearing on the matrix element level is provided by Ref.~\cite{NieMiera:2025inn} for similar kaon gluon data.
A key comparison here is that the ratio of renormalized Wilson3 smearing is similar to about 10 steps of HYP smearing.}

%\BG{Description of the $m_0 +\delta m$ fit}
With the zero momentum matrix elements, $h^\text{B}(z,0)$, and the Wilson coefficient defined in Eq.~\ref{eq:wil-coef}, we may fit $\delta m + m_0$ using the form in Eq.~\ref{eq:dm_fit}.
%apply hybrid renormalization with $\mu=2.0$~GeV to the matrix elements with non-zero momentum.
The $a\approx 0.12$~fm lattice spacing is too coarse to capture where $\ln{\left[\mathcal{H}(z,\mu) /h^{\text{B}}(z,0) \right]}$ behaves linearly in the small-$z$ region, so we interpolate the $h^\text{B}(z,0)$.
We fit the interpolated data in the set $\{z-0.2, z, z+0.2\}$ in units of fm, varying $z$.
\UPDATE{We find that the results are highly consistent for any order of interpolation between 3 and 8.}
We show these results for the $\delta m + m_0$ versus $z$ for each hadron and each smearing, in Fig.~\ref{fig:m0_vs_z} with statistical errors only.
We do not consider scale variation here.
The general behavior of $\delta m + m_0$ is very reasonable here for all cases.
The fits are unstable in the small-$z$ region due to discretization effects, then there is a nice plateau of stability, before we see the hint of divergence due to perturbation theory breaking down as $z$ becomes larger.
As seen by comparing the left and right plots, the amount of smearing can have a significant effect on the value of $\delta m + m_0$.
Focusing on the higher precision Wilson3 data on the left plot, we see that $\delta m + m_0$ has very little dependence on the external state, with hadrons at the same pion mass being statistically indistinguishable and the same hadrons at different pion masses having agreement well within about $2\sigma$.
These results are in contrast to those from the other two operators explored in Fig. 11 of Ref.~\cite{Good:2024iur}.
For those operators, the fitted $\delta m + m_0$ values depend highly on the external states and do not appear to level off as nicely.
This adds even further to the list of reasons why the operator in this work is preferable to the other two.
We choose $\text{min}_z(\delta m + m_0)$ as the value to use for hybrid renormalization moving forward for each hadron and smearing, as these happen to be where the curves are the flattest.
We note that, for Wilson-3, these values are between $0.35$ and $0.4$ GeV, which is slightly smaller than the value $0.5$ GeV, which we estimated in our previous paper without knowing the Wilson coefficients.

\begin{figure}[t]
    \centering
    \includegraphics[width=\linewidth]{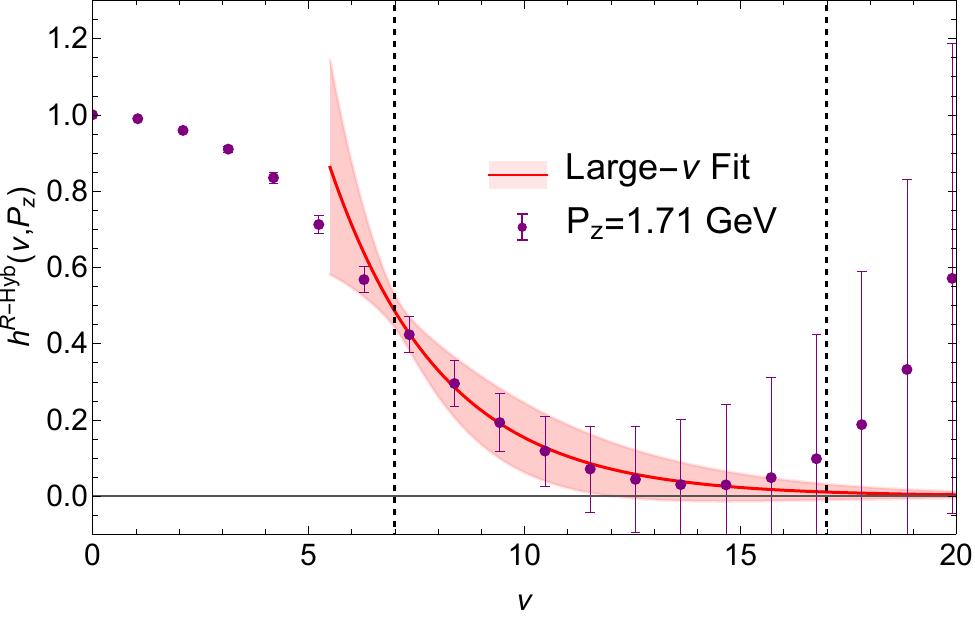}
    \caption{Large-$\nu$ extrapolation of the hybrid renormalized $P_z = 1.71$ GeV matrix elements \UPDATE{with $M_\pi \approx 690$ MeV and Wilson3 smearing}, using the data from $z \in [7a, 16a]$, the extent of which is shown by the two dashed, vertical lines.}
    \label{fig:O0IB_large-nu}
\end{figure}

%\BG{Hybrid renormalized matrix elements}
We plot the hybrid renormalized matrix elements using the fitted $\delta m + m_0$ values from the zero momentum matrix elements for the light and strange nucleons for the Wilson-3 and HYP5 smearing along with the nucleon matrix elements reconstructed from the CT18 NNLO gluon PDF~\cite{Hou:2019efy} in Fig.~\ref{fig:hybrid-lat-pheno} \UPDATE{to provide a qualitative example of matrix elements from a physically reasonable PDF.}
\UPDATE{These matrix elements are computed by using the CT18 gluon PDF on the righthand side of Eq.~\ref{eq:mtcheq} to obtain the quasi-PDF, then Fourier transforming back to position space.}
We use a value of $z_s = 3a \approx 0.36$~fm. 
Looking at the lattice data first, as one would expect from increased gauged link smearing, we see that the Wilson-3 matrix elements seem to fall off more slowly than the HYP5 matrix elements.
This suggests that the Wilson-3 smearing is affecting the physics, especially when we compare with the phenomenological matrix elements.
We see that the low momentum matrix elements seem to fall off much faster and disagree more significantly with the phenomenological matrix elements, with the largest momentum matrix elements beginning to converge more quickly in the strange nucleon case.
We are not overly concerned with the lowest momenta data because we intend to apply LaMET matching only to large-momentum data, but this may be interesting to explore further to better understand the systematics.
It seems that the HYP5 strange nucleon is in the best agreement with the phenomenological results, while the ``most physical'' data should be from the light nucleon for HYP5.
This could be some canceling of various systematic effects, which will all require further exploration.
One further curiosity is that there is still a bump near $\nu_s = z_sP_z$ in the phenomenological matrix elements for the lowest momentum that was seen in our previous results for the other two gluon operators in Fig. 14 of Ref.~\cite{Good:2024iur}; however, it is much smaller and not very present in the larger momentum results at all.
This is due to the differences in the matching for the different operators and has nothing to do with the lattice value of $\delta m + m_0$.

\begin{figure}
    \centering
    \includegraphics[width=\linewidth]{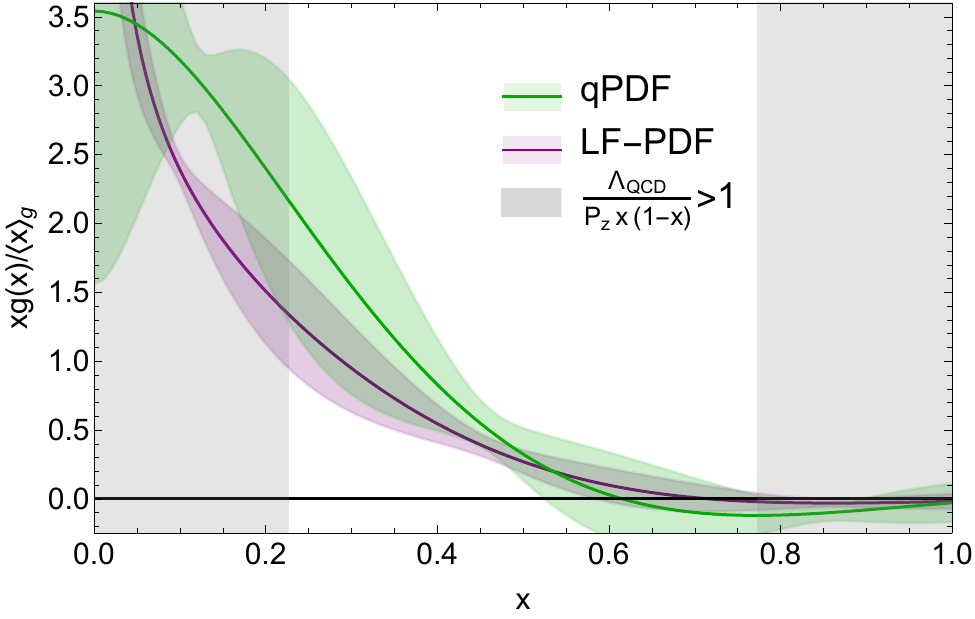}
    \caption{Quasi-PDF (green) from the Fourier transform of the hybrid renormalized $P_z = 1.71$ GeV matrix elements with $M_\pi \approx 310$ MeV and HYP5 smearing and the light-front (LF-)PDF (purple) after matching the quasi-PDF (qPDF) to the light cone.
    The gray bands represent the region where LaMET matching begins to break down.}
    \label{fig:qPDF-PDF-comp}
\end{figure}

\begin{figure*}[ht!]
    \centering
    \includegraphics[width=.45\linewidth]{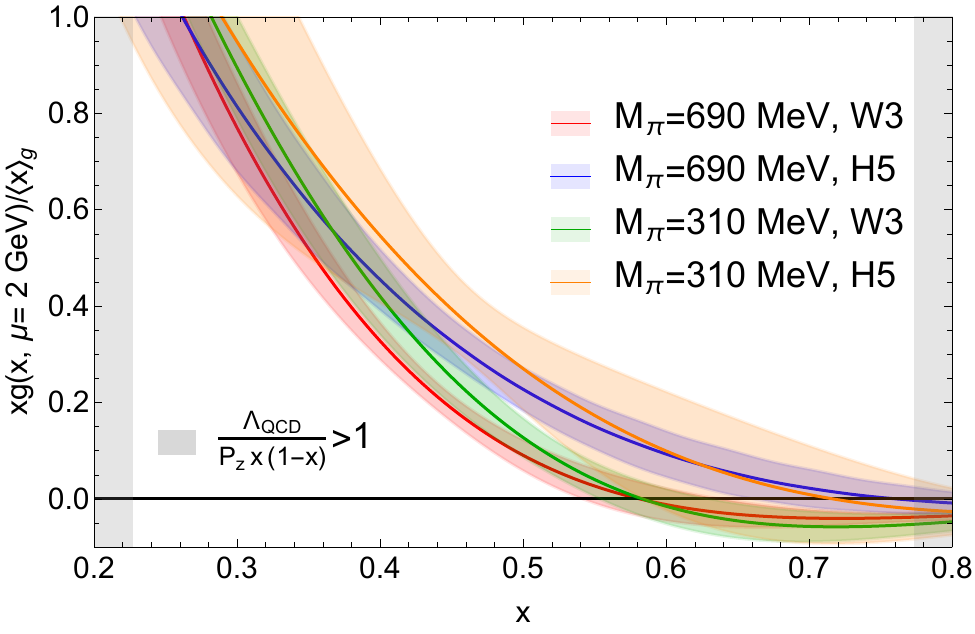}
    \includegraphics[width=.45\linewidth]{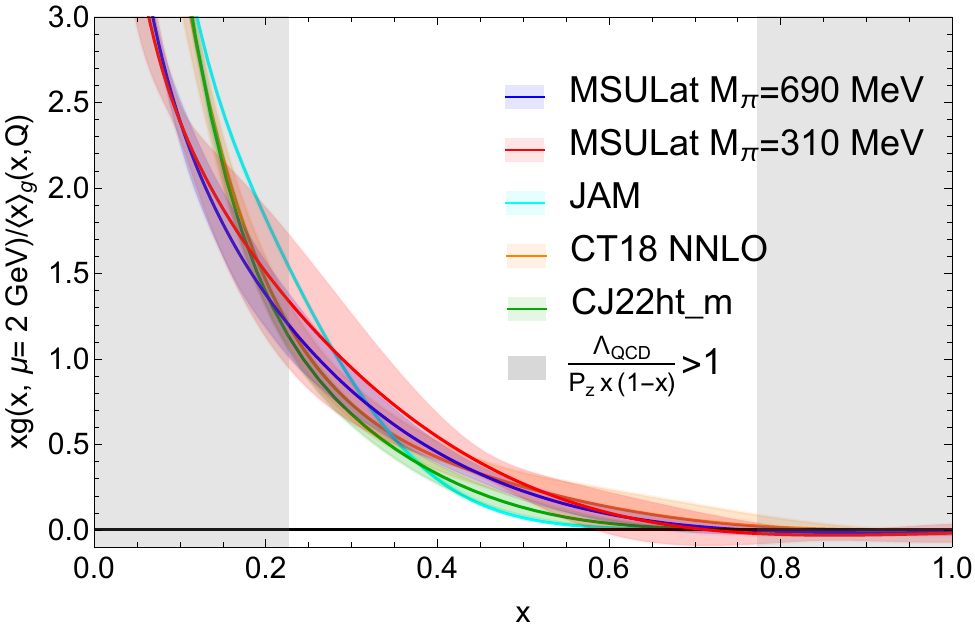}
    \caption{ (Left) the MSULat nucleon gluon PDFs at $P_z = 1.71$ for the different pion masses and smearings all compared. 
    (Right) The MSULat PDF  from large momentum effective theory at $P_z = 1.71$ GeV at $M_\pi \approx 690$ (blue) and $310$ (red) MeV with HYP5 smearing compared to the CT18~\cite{Hou:2019efy} (orange), JAM24~\cite{Anderson:2024evk} (cyan), and CJ22 with multiplicative higher twist corrections~\cite{Cerutti:2025yji} (green) gluon PDFs.
    In both plots, the gray bands represent the region where LaMET matching begins to break down.
    }
    \label{fig:PDF-GF-comp}
\end{figure*}

%\BG{Large-$\nu$ extrapolation}
Thanks to the extremely high statistics, we have reasonable signal at large enough distances and high enough momentum to apply a large-$\nu$ extrapolation to do a Fourier transformation.
We follow previous work and use an ansatz of the form~\cite{Ji:2020brr}
\begin{equation}\label{eq:large-nu-form}
    h^\text{R}(z,P_z) \approx A\frac{e^{-m\nu}}{|\nu|^d}
\end{equation}
to fit the large-$\nu$ data, where $A$, $m$ and $d$ are fitted parameters.
We use this form to fit our nucleon data at $P_z = 1.71$~GeV data for each smearing and pion mass.
\UPDATE{This fit form has strong theoretical motivation described in the supplemental material of Ref.~\cite{Gao:2021dbh}, which allows us to replace the noisy long-distance data with the well constrained, physically motivated extrapolation.
This has been compared to standard lattice methods of using theoretically motivated fit forms to extract information from limited data, justifying the truncation of noisy data.
One such example is extracting matrix elements from lattice correlator data, which is described in our earlier paper~\cite{Good:2024iur}.}
The fit window for the extrapolation that we used in our previous work~\cite{Good:2024iur} was overly conservative, as we can fit data as low as $z = 6a = 0.72$ fm, and still achieve consistent results, which is explored further in Appendix~\ref{sec:app_large-nu-comp}.
\UPDATE{The results imply that there is systematic uncertainty associated with the fit range.
We do not explore this systematic in detail in this first proof of principle calculation.}
We use fit windows that start at $6a$ for the HYP5 data, and $7a$ for the Wilson3 data.
We find again that it is still difficult to separate the algebraic and exponential decay at this level of precision, which causes a very large amount of fluctuation in the fit parameters, but this fluctuation of the individual parameters cancels and allows us to obtain a reasonable fit and extrapolation.

We plot the results of this fit for the Wilson3 strange nucleon in Fig.~\ref{fig:O0IB_large-nu}, as an example of our most precise data.
We see qualitatively that the fit agrees well with the data and decays nicely to zero while the largest distance data diverge \UPDATE{due to the increasing UV fluctuations and finite volume effects introduced at such large distances}.
Inside the fit range, the parameterization has error similar to the data, before shrinking exponentially compared to the exponentially growing error in the data.
This is consistent for the other datasets, which are shown in Appendix~\ref{sec:app_large-nu}.
With these two features, we can reliably perform a Fourier transformation on the data.

%\BG{quasi-PDF and PDF}
We can connect an interpolation of the short distance lattice matrix elements and the large-$\nu$ extrapolation at longer distances to compute the Fourier transformation of the matrix elements, which defines the quasi-PDF.
We apply the light-front matching defined in Eqs.~\ref{eq:mtcheq} and \ref{eq:kernel} to obtain the light-front PDF.
As an example of the other end of the spectrum of our statistical precision, Fig.~\ref{fig:qPDF-PDF-comp} shows the results for the quasi- and light-front-PDFs for the light nucleon with HYP5 smearing, with the regions with $\frac{\Lambda_\text{QCD}}{P_zx(1-x)} \gtrsim 1$, where LaMET matching breaks down, grayed out.
\UPDATE{We use $\Lambda_\text{QCD} = 300$ MeV as a conservative estimate.}
Focusing first on the quasi-PDF in green, we see that there is some negativity in the quasi-PDF which was also seen in our previous estimation in Ref.~\cite{Good:2024iur}.
We also see pinch points in the errors on the quasi-PDF, which is likely due to the fairly large range in the decay rates found in our large-$\nu$ extrapolation.
This is most significant for the HYP5 light nucleon, which we show here.
A nice feature is that the quasi-PDF is very well convergent to 0 as $x \rightarrow 1$.
Comparing the light-front- and quasi-PDFs, we can see that the matching has a significant effect on the PDFs. 
This seems to be larger than what is often seen in the quark PDFs, but this can be partially accounted for due to the difference in the factors $C_F = 4/3$ and $C_A = 3$ in front of the NLO term in the matching for the quark and gluon PDFs, respectively.
The negative portion of the quasi-PDF is mostly removed by the matching, with the light-front PDF having a slightly negative central value from $x \gtrsim 0.75$, which is mostly within $1\sigma$ of zero.
Another striking effect of the matching is that the errors seem to be significantly reduced.
It could be that the oscillation in the errors of the quasi-PDF undergoes some level of destructive interference through the matching or that the matching is simply very robust to the statistical fluctuations in the data.
It will be interesting to explore this further as the data becomes more precise.

%\BG{Comparison with pheno. results?}
To get a picture of the systematics from smearing and pion masses, we show all of our nucleon gluon PDFs from the $P_z = 1.71$~GeV lattice data at the two pion masses and two smearings with only statistical fluctuations on the left side of Fig.~\ref{fig:PDF-GF-comp}.
The Wilson3 smearing data produces PDFs that are more negative at $x \gtrsim 0.6$, but within about 1 or 2$\sigma$ of zero.
At $x \approx 0.5$, where LaMET theory is most reliable, the $M_\pi \approx 690$ MeV PDFs are around $3\sigma$ from each other, which demonstrates that there are significant effects on the physics caused by the very large Wilson3 smearing.
\UPDATE{In particular, this effect in the large-$x$ may be coming from the effects of smearing on the short-distance physics or perhaps imperfect renormalization, which has not been proven analytically for smeared lattice studies.}
Overall, we believe that the HYP5 PDFs are more reliable than the Wilson3 PDFs.
The PDFs at the two different pion masses for each smearing are within one sigma of each other, with those at the smaller pion mass having larger statistical fluctuations, as one would expect.
There is still more work to be done in understanding how the systematic effects emerge in the final PDFs, but this is not the focus of this first study.
\UPDATE{It would be interesting to explore renormalizability in the continuum theory with Wilson flow.}

We compare our HYP5 PDFs at each pion mass to the CT18~\cite{Hou:2019efy}, JAM24~\cite{Anderson:2024evk}, and the CJ22 multiplicative higher twist effect~\cite{Cerutti:2025yji} gluon PDFs, which are fairly different global fits from each other, in Fig.~\ref{fig:PDF-GF-comp}.
We see that the HYP5 lattice PDFs actually agree very well with the CT18 across the entire region, with the strange nucleon agreeing particularly well, which is expected from the agreement seen in position space in Fig.~\ref{fig:hybrid-lat-pheno}.
We see that on the lower and upper side of the region that we trust, we see 1$\sigma$ agreement with the CJ22 PDFs, while it seems that the agreement is lesser around $x\approx 0.5$ due to the pinch point in the lattice error bars here.
We see some overlap with the JAM24 PDF at several different points in the trustworthy region, with particular agreement with the lighter pion mass, partially due to the larger errors.
Overall, our PDF falls within the range of global fits, preferring a slightly larger gluon PDF around $x\approx 0.5$ than most global fits.
\UPDATE{This provides a loose validation that the methodology can provide qualitatively reasonable gluon PDF; however, there are many systematic effects to be understood.
At this level, it is not abundantly clear whether there is some cancellation of systematic effects causing the agreement.}
It will be interesting to see what occurs as noise reduction techniques reduce the error on the lighter pion mass results and other lattice systematics are taken into consideration, such as physical continuum extrapolations.
Finer lattice spacings will be needed to go to higher momenta, as well as to increase the reliable region of our PDFs. 

%%%%%%%%%%%%%%%%%%%%%%%%%%%%%%%%%%%%%%%%%%%%%%%%%%%%%%%%%%%%%%%%%%%%%%%%%%%
\section{Conclusion and Outlook}
\label{sec:conclusion}

We have presented the first nucleon gluon PDF calculated through LaMET with hybrid-ratio renormalization with at two heavier than physical pion masses, $690$ and $310$ MeV, using two gauge link smearing techniques, Wilson flow with a flow time of $3a^2$ and 5 steps of hypercubic smearing.
Based on the prior work toward a LaMET gluon PDF~\cite{Good:2024iur}, we focused on a single gluon operator $O^{(3)}$ defined in Eq.~6 of Ref.~\cite{Good:2024iur}, which has been used in unpolarized gluon studies using pseudo-PDF method~\cite{Fan:2020cpa,Fan:2021bcr,HadStruc:2021wmh,Salas-Chavira:2021wui,Fan:2022kcb,Delmar:2023agv,Good:2023ecp}.
We computed the one-loop Wilson coefficient needed for hybrid renormalization for this operator and the corresponding one-loop matching kernel.
Using these Wilson coefficients, we found $\delta m+m_0$ to be consistent between the pion and nucleon, whether at the light or strange mass, in contrast with previous results for the $O^{(1)}$ and $O^{(2)}$ operators in Ref.~\cite{Good:2024iur}.
We computed the nucleon gluon PDFs for each pion mass and smearing technique with hadron boost momentum $P_z = 1.71$~GeV.
We found that the Wilson3 smearing significantly affects the physics, producing PDFs that are negative around $x \gtrsim 0.6$, while the noisier HYP5 PDFs do not fall significantly below zero.
We compared the HYP5 nucleon gluon PDFs at both pion masses with from the CT18~\cite{Hou:2019efy}, JAM24~\cite{Anderson:2024evk}, and CJ22~\cite{Cerutti:2025yji} global fits as a small sample.
However, we find good agreement with the CT18 PDF across the full region, and less agreement with CT22 and JAM24 around $x\approx 0.5$, which shows that our PDF falls comfortably within the range of those found in global fits.
\UPDATE{The final error bars on the lattice PDFs do not estimate systematics from the extrapolations to complete the Fourier transform, the demonstration of which was the central challenge in these first proof of principle results.
Bayesian and machine learning techniques~\cite{Chowdhury:2024ymm,Dutrieux:2024rem,Dutrieux:2025jed} could be transferred to expand on the simple extrapolation model to give a more rigorous error quantification.}
% FUTURE
This study demonstrates a strong proof-of-principle for using large momentum effective theory to calculate gluon PDFs, which was long thought to be out of reach.
We believe that this work provides strong motivation to use time and resources to replace the large pion mass and smearing with noise-reduction techniques with fewer physics effects to obtain more reliable gluon PDFs from LaMET.
We plan to improve the current calculation by using self-renormalization scheme~\cite{LatticePartonLPC:2021gpi} with zero-momentum boosted matrix elements with multiple lattice spacings.
\UPDATE{This will allow us to better understand discretization and smearing effects and improve the systematics on the gluon PDFs by taking the continuum-physical limit.}
As the errors become smaller, there will also be a need to understand the effects of scale variation in the gluon PDFs from LaMET.

%%%%%%%%%%%%%%%%%%%%%%%%%%%%%%%%%%%%%%%%%%%%%%%%%%%%%%%%%%%%%%%%%%%%%%%%%%%

\section*{Acknowledgments}
We thank the MILC Collaboration for sharing the lattices used to perform this study.
The LQCD calculations were performed using the Chroma software suite~\cite{Edwards:2004sx}.
This research used resources of the National Energy Research Scientific Computing Center, a DOE Office of Science User Facility supported by the Office of Science of the U.S. Department of Energy under Contract No. DE-AC02-05CH11231 through ERCAP;
facilities of the USQCD Collaboration, which are funded by the Office of Science of the U.S. Department of Energy,
and supported in part by Michigan State University through computational resources provided by the Institute for Cyber-Enabled Research (iCER).
The work of WG is partially supported by U.S. Department of Energy, Office of Science, under grant DE-SC0024053 ``High Energy Physics Computing Traineeship for Lattice Gauge Theory'' and by the US National Science Foundation under grant PHY~2209424.
FY is supported by the U.S. Department of Energy, Office of Science, Office of Nuclear Physics through Contract No. DE-SC0012704, and within the framework of Scientific Discovery through Advanced Computing (SciDAC) award Fundamental Nuclear Physics at the Exascale and Beyond.
HL is partially supported by the US National Science Foundation under grant PHY~2209424.

%%%%%%%%%%%%%%%%%%%%%%%%%%%%%%%%%%%%%%%%%%%%%%%%%%%%%%%%%%%%%%%%%%%%%%%%%%%
%\vfill\null
%\vfill\null
\begin{figure}[!th]
    \centering
    \includegraphics[width=\linewidth]{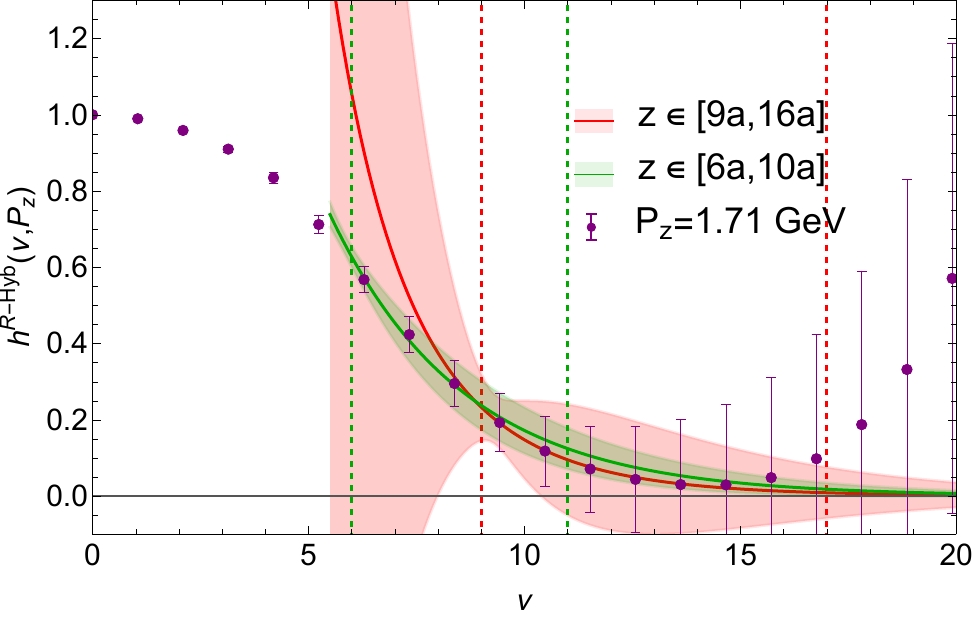}
    \caption{Two large-$\nu$ extrapolations of the hybrid renormalized $P_z = 1.71$ GeV matrix elements at $M_\pi \approx 690$ MeV, with Wilson3 smearing.
    The two extrapolations are fit using the data $z \in [6a, 10a]$ (green) and $[9a,16a]$ (red).
    The range of each is shown by the dashed, vertical lines in each plot. }
    \label{fig:ME_large_nu_comp}
\end{figure}

\begin{figure}[!th]
    \centering
    \includegraphics[width=\linewidth]{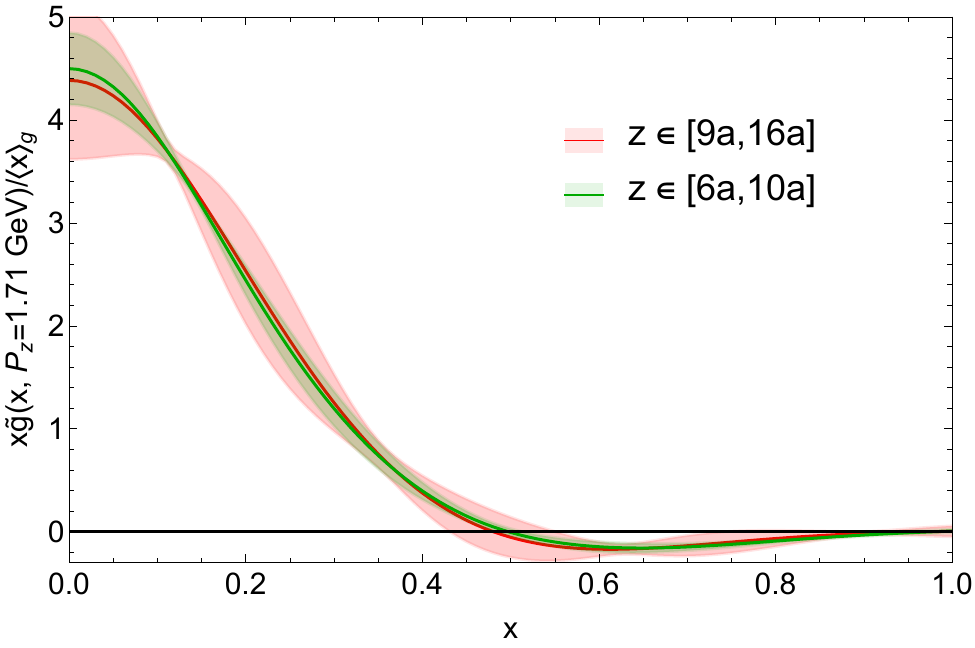}
    \caption{The quasi-PDFs produced from the $z \in [6a, 10a]$ (green) and $[9a,16a]$ (red) large-$\nu$ extrapolations of the $M_\pi \approx 690$ MeV data with Wilson3 smearing.}
    \label{fig:qPDF_large_nu_comp}
\end{figure}

\begin{appendices}

\begin{figure}[H]
    \centering
    \includegraphics[width=\linewidth]{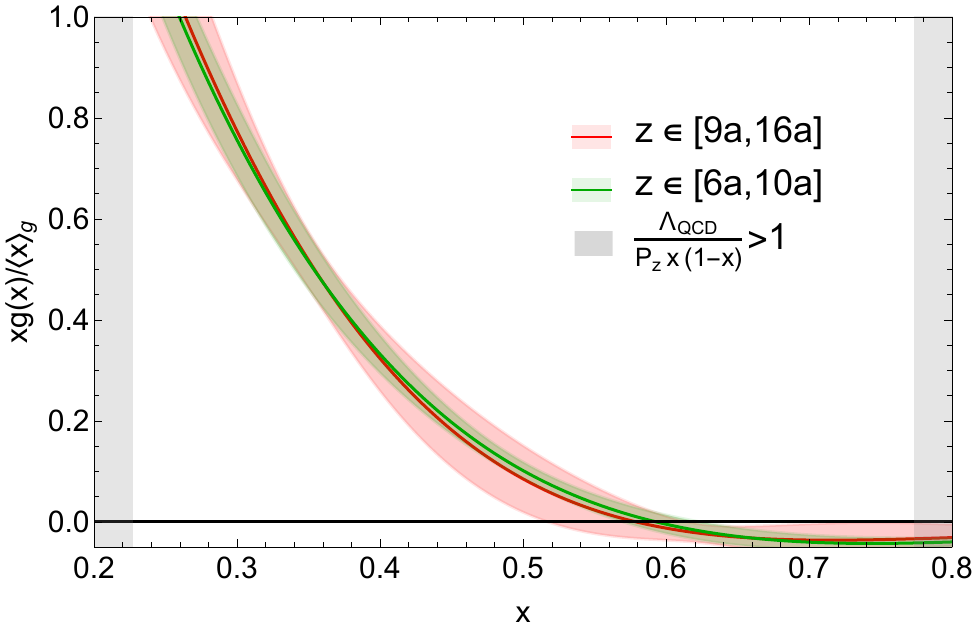}
    \caption{The light-front PDFs produced from the $z \in [6a, 10a]$ (green) and $[9a,16a]$ (red) large-$\nu$ extrapolations of the $M_\pi \approx 690$ MeV data with Wilson3 smearing.}
    \label{fig:PDF_large_nu_comp}
\end{figure}

\section{Extrapolation fit range effects}\label{sec:app_large-nu-comp}
In this section, we compare two different large-$\nu$ the $M_\pi \approx 690$ MeV Wilson3 data, the cleanest data, to demonstrate the stability of the results.
In Fig.~\ref{fig:ME_large_nu_comp}, we show the extrapolations from the data within the ranges $z \in [6a, 10a]$ (green) and $[9a,16a]$ (red).
We observe that the two extrapolations are both comfortably within 1$\sigma$ of each other and the data from $\nu \approx 6-20$; however, the errors are quite different.
The extrapolation to the larger distance data gives much larger error, while the shorter distance data can be used to much more strongly constrain the extrapolation.
In the main text, we use an extrapolation from $z \in [7a, 16a]$, to make use of the error reduction provided by the shorter distance data, while also fitting better to the large $\nu$.

Moving forward, we compute the Fourier transform to obtain the quasi-PDFs for each extrapolation, which we show in Fig.~\ref{fig:qPDF_large_nu_comp}.
We see that the central values are in very good agreement.
Through the majority of the range, the smaller $z \in [6a,10a]$ errors encompass the mean of the $z \in [9a,16a]$ quasi-PDF.
Again, the larger-distance extrapolation has much larger error bars.
Despite the stronger constraint on the extrapolation from the $z \in [6a, 10a]$ data, there is oscillation and pinch points in the errors in both quasi-PDFs.

Finally, in Fig.~\ref{fig:PDF_large_nu_comp}, we show the light-front PDFs produced from the two extrapolations.
Similar to the quasi-PDFs, the central values are consistently within 1$\sigma_{z\in[6a,10a]}$ of each other; however, the errors are quite different.
These results demonstrate that the central value of the PDFs remain fairly stable despite the difference large-$\nu$ extrapolations, while error quantification may need to be explored more thoroughly.
There are new techniques that could be used to supplement the simple $e^{-m\nu}/\nu^d$ extrapolation, including Bayesian and machine learning techniques, some of which have been developed for the pseudo-PDF methodology, but could be transferred to LaMET~\cite{Chowdhury:2024ymm,Dutrieux:2024rem,Dutrieux:2025jed}.
This should be explored further in future work.

\section{Large-$\nu$ extrapolations}\label{sec:app_large-nu}
\begin{figure*}[h!]
    \centering
    \includegraphics[width=.45\linewidth]{Figs/O0IBncls_W3_large-nu.pdf}
    \includegraphics[width=.45\linewidth]{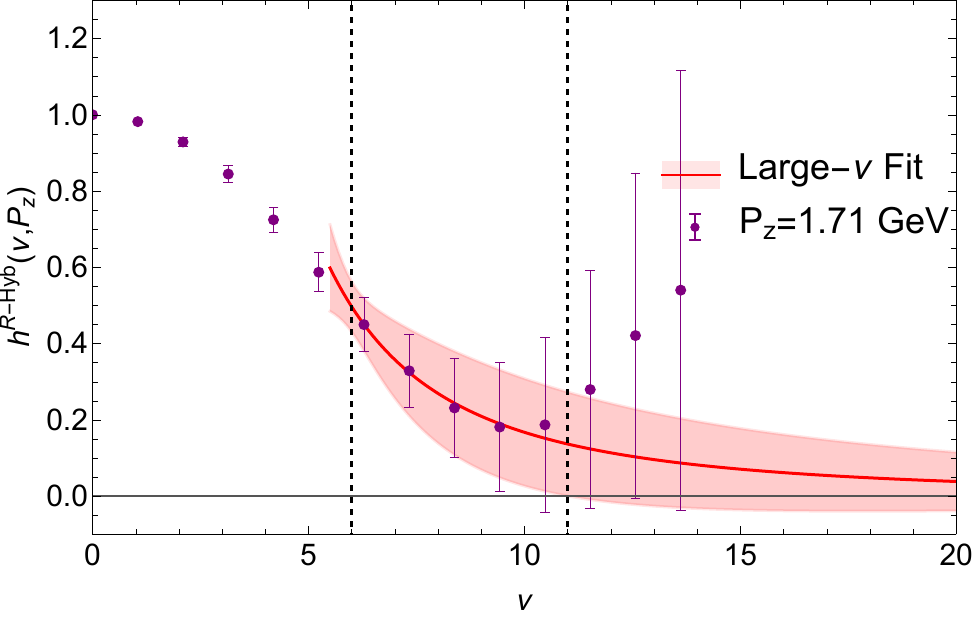}
    \includegraphics[width=.45\linewidth]{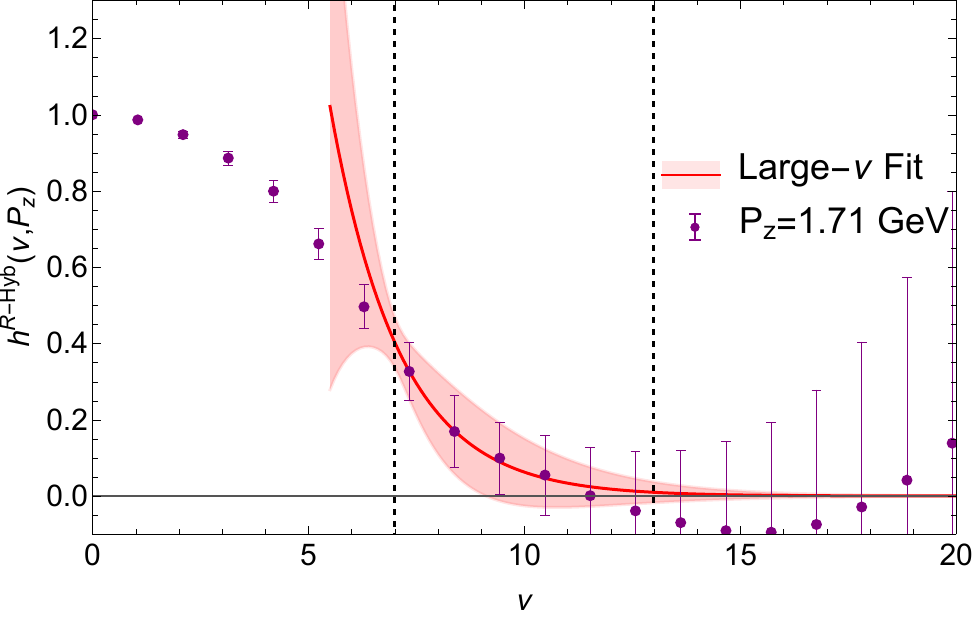}
    \includegraphics[width=.45\linewidth]{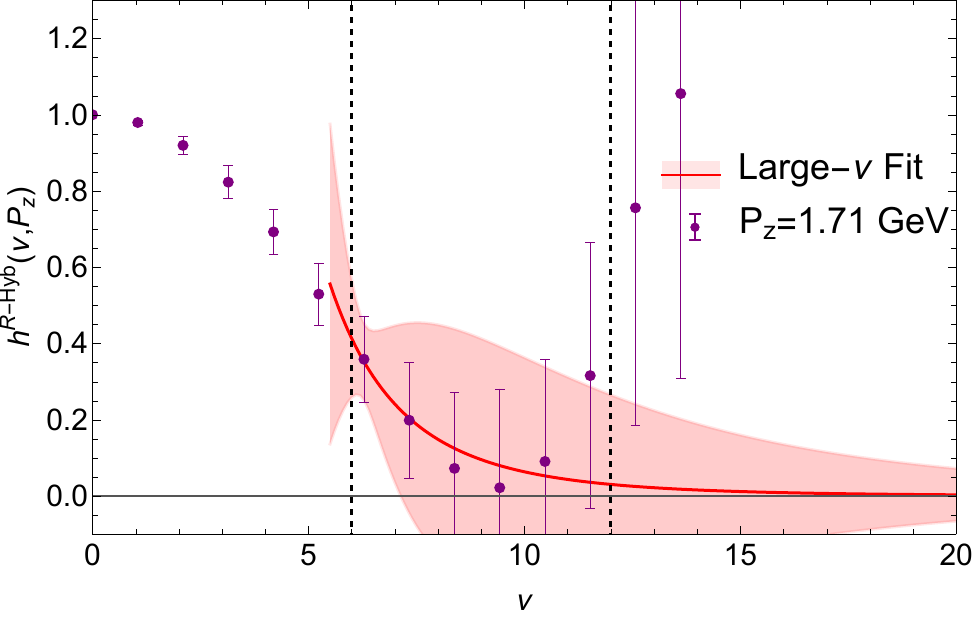}
    \caption{Large-$\nu$ extrapolation of the hybrid renormalized $P_z = 1.71$ GeV matrix elements at $M_\pi \approx 690$ (top) and $310$ (bottom) MeV  with Wilson3 (left) and HYP5 (right) smearing.
    The extrapolations are fit using the data between the two dashed, vertical lines in each plot.}
    \label{fig:O0IB_large-nu-all}
\end{figure*}

For completeness, we show the additional large-$\nu$ extrapolations in Fig.~\ref{fig:O0IB_large-nu-all} in this section.
The fit ranges used for the results in the main text are $[7a,16a]$ and $[7a,12a]$ ($[6a,10a]$ and $[6a,11a]$) for the strange and light Wilson3 (HYP5) nucleon data respectively.
In the previous appendix section, we showed that fit ranges starting as high as $z = 6a$ are valid choices.
We see that the Wilson3 data is much better constrained at long distances than the HYP5. 
This results in the HYP5 extrapolations fall off more slowly those for the Wilson3, but we do not expect this long distance behavior to change drastically change the physics in the trustworthy $x \sim 0.2-0.8$ range.

\end{appendices}

\bibliography{refs}

\end{document}